\documentclass[11pt, reqno, letter]{amsart}
\usepackage{setspace}
\usepackage[foot]{amsaddr}
\usepackage{tikz-qtree}

\newtheorem{theorem}{Theorem}

\newcommand{\var}{\text{var}}
 
\allowdisplaybreaks

\title[Discovering Effect Modification in Air Pollution Studies]{Discovering Effect Modification and Randomization Inference in Air Pollution Studies}
\author{Kwonsang Lee$^{1}$, Dylan S. Small$^{2}$, and Francesca Dominici$^{1}$}
\address{$^{1}$ Department of Biostatistics, Harvard School of Public Health}
\address{$^{2}$ Department of Statistics, University of Pennsylvania}

\begin{document}

\begin{abstract}
Studies have shown that exposure to air pollution, even at low levels, significantly increases mortality. As regulatory actions are becoming prohibitively expensive, robust evidence to guide the development of targeted interventions to reduce air pollution exposure is needed. In this paper, we introduce a novel statistical method that splits the data into two subsamples: (a) Using the first subsample, we consider a data-driven search for {\it de novo} discovery of subgroups that could have exposure effects that differ from the population mean; and then (b) using the second subsample, we quantify evidence of effect modification among the subgroups with nonparametric randomization-based tests. We also develop a sensitivity analysis method to assess the robustness of the conclusions to unmeasured confounding bias. Via simulation studies and theoretical arguments, we demonstrate that since we discover the subgroups in the first subsample, hypothesis testing on the second subsample can focus on theses subgroups only, thus substantially increasing the statistical power of the test. We apply our method to the data of 1,612,414 Medicare beneficiaries in New England region in the United States for the period 2000 to 2006. We find that seniors aged between 81-85 with low income and seniors aged above 85 have statistically significant higher causal effects of exposure to PM$_{2.5}$ on 5-year mortality rate compared to the population mean.  
\end{abstract}

\keywords{Causal effect, Causal inference, Fine Particulate Matter, Observational study, Recursive partitioning, Sample split}



\maketitle

\doublespace

\section{Introduction}

Air pollution is a major environmental risk to health. Over the past few decades, researchers have estimated the  association between air pollution exposure and a wide range of health outcomes from respiratory diseases to death (Dockery et al. 1993; Samet et al. 2000;  Dominici et al. 2006; Loomis et al. 2013; Di et al. 2017; Makar et al. 2017). Recently, Di et al. (2017) reported statistically significant evidence of increased mortality risk associated with long term exposure to PM$_{2.5}$ even when these levels are always below the  current national ambient air quality standards. The World Health Organization (WHO)'s International Agency for Research on Cancer (IARC) has concluded that exposure to outdoor air pollution, especially, fine particulate matter (PM$_{2.5}$), is carcinogenic to humans (Loomis et al. 2013). Major strides have been made to prove adverse causal effect of outdoor air pollution, and regulations should be followed to decrease air pollution and promote public health. As the air quality regulation costs become expensive, robust evidence of more targeted regulatory actions is required for determining the appropriate allocation of regulatory efforts and resources. Our goal in this paper is to answer the question ``What subgroups have different effects from the overall population mean?'' Answering to this question has an influential impact on regulations. For example, the Clean Air Act requires the U.S. Environmental Protection Agency (EPA) to set the national air quality to protect subgroups that are sensitive to air pollution. Therefore, it is important and informative to discover potentially sensitive subgroups in order to establish regulatory programs such as setting more stringent national standards for air pollutants. There have been several epidemiological studies on discovering sensitive subgroups and estimating the effects of air pollution in these subgroups, but most of the studies have two main limitations; (a) the existing approaches rely on standard regression approaches for confounding adjustment, without accurate balance checking and systematic sensitivity analyses for unmeasured confounding; and (b) variables that are suspected to be potential effect modifiers are selected a priori before analysis. In this paper, we propose to overcome these limitations in the framework of causal inference by letting the data to discover vulnerable subgroups and using nonparametrical tests with a sensitivity analysis for unmeasured confounding.

We conduct an observational study of all Medicare beneficiary residing in New England region in the United States that contains six states (Maine, Vermont, New Hampshire, Massachusetts, Rhode Island, and Connecticut) between 2000-2006. Our goal is to discover subgroups that have statistically significantly different causal effects of long-term exposure to PM$_{2.5}$ on mortality from the population average. The main challenge in observational studies is to remove confounding bias. Matching is a simple and transparent way to adjust for biases due to measured confounders (Stuart 2010). Roughly speaking, for each treatment subject, matching produces a strata by placing in controls who are similar to the treated. Once the produced matched sets or pairs pass diagnostics such as the overall covariate balance that checks how similar the treated and control groups are with respect to their measured potential confounders, further inferences can be allowed to proceed. Under the assumption of no unmeasured confounding bias, inferences can be made by treating matched sets as stratified randomized experiments (Rosenbaum 2002a; Hansen 2004). To test whether there is effect modification, randomization-based tests for the null hypothesis of a constant treatment effect across the population are considered. Randomization inference does not use any model assumptions, therefore randomization-based tests are nonparametric. The assumption of random assignment of treatment is used to provide a reasoned basis for inference. See Rosenbaum (2002b) for more detailed discussion of randomization inference in observational studies.

In this paper, we introduce a novel approach to {\it de novo} discovery of effect modification followed by confirmatory hypothesis testing. More specifically, we split the sample into two parts. In the first subsample, we discover ``promising'' subgroups with heterogeneous treatment effects. In the second subsample, we develop randomization-based hypothesis tests to confirm evidence of effect modification. In the discovery step, we apply two machine learning algorithms to uncover heterogeneous structures of treatment effects: (a) classification and regression tree (CART) method proposed by Breiman et al. (1984); and (b) Causal tree (CT) method proposed by Athey and Imbens (2016) using different criteria for constructing partitions. In the confirmation step, we integrate  the discovered tree structure into a randomization-based testing framework, to provide evidence as whether subgroup exposure effects are statistically significantly different than the population average.

In observational studies, the assumption of no unmeasured confounding bias is often too restrictive. Even when this assumption is violated, randomization inference can be generalized to assess the robustness of the results to unmeasured confounding bias. In this paper, we consider a sensitivity analysis to characterize how extensive an unmeasured confounding bias has to be to alter our conclusions regarding evidence of effect modification.


Our proposed method for discovery and testing of effect modification is innovative and has several desirable features. First, in the confirmatory phase, our new approach is designed to test the deviance of subgroup treatment effects from the population mean, not from the null effect as done in the existing literature (Hsu et al. 2013; Lee et al. 2017). Rejecting these hypotheses for certain subgroups can provide statistical evidence and practical guidance to look more closely at the subgroups for future researchers. Second, our method considers sensitivity to unmeasured biases for the estimation of the population mean that is a main concern in observational studies. There are several works for estimating average treatment effects; For example, Wager and Athey (2017) propose a method based on random forests to estimate the covariate-specific treatment effects and Su et al. (2009) use recursive partitioning to estimate treatment effects across subpopulations. However, most of the works do not consider sensitivity analyses to unmeasured confounding. Our proposed method accounts for unmeasured biases in estimating the population mean and also in testing effect modification in subgroups. Third, in high dimensional settings, some important variables in a tree can be discovered by using the first subsample obtained from the sample-splitting approach. This will increase statistical power to detect effect modification even if the size of the testing subsample is reduced. Finally, our method provides results that can be directly used for regulatory policy. Since a tree is produced with highlighted subgroups having significantly different treatment effects from the population mean, such tree structures can be easily explained to non-experts. 

The outline of the rest of this paper is as follows. In Section 2, we review observational studies, matching and sensitivity analyses. In Section 3, we describe our proposed method for continuous outcomes and for binary outcomes. Here we also review and compare other methods for de novo discovery of effect modification. In Section 4, we illustrate the performance of our method in several simulated situations. We apply our method to the air pollution data of Medicare beneficiaries in Section 5. Section 6 concludes with a discussion. 

\section{Notation and review of observational studies}
\label{s:notation}

\subsection{Notation For Stratified Randomized Experiments }
\label{ss:notation}

Suppose that there are $G$ groups of matched sets. For group $g$, there are $I_g$, $g=1, \ldots, G$, matched sets and for set $i$ in group $g$ and $n_{gi}$, $i=1, \ldots, I_g$, individuals. For each set $i$ in group $g$, $m_{gi}$ individuals receive the treatment and $n_{gi}-m_{gi}$ individuals receive the control with $\min\{m_{gi}, n_{gi}-m_{gi}\} = 1$. For simplicity, we assume $m_{gi}=1$, which means that each $gi$ has only one treated individual. If individual $gij$ receives a treatment, we denote $Z_{gij}=1$ otherwise $Z_{gij}=0$. For each $gi$, $\sum_{j=1}^{n_{gi}} Z_{gij}= 1$. Under the potential outcome framework with binary treatment, each individual has two potential outcomes; one is under treatment, $r_{Tgij}$ and the other is under control, $r_{Cgij}$. Only one of the two potential outcomes can be observed according to treatment assignment $Z_{gij}$, thus the individual treatment effect, $r_{Tgij} - r_{Cgij}$ cannot be observed; see Neyman (1923) and Rubin (1974). This individual exhibits the observed response $R_{gij} = r_{Tgij} Z_{gij} + r_{Cgij}(1-Z_{gij})$

Let $\mathcal{F} = \{(r_{Tgij}, r_{Cgij}, \mathbf{x}_{gij}), g=1, \ldots, G, i=1, \ldots, I_g, j=1, 2 \}$ where $\mathbf{x}_{gij}$ denotes observed covariates, and $\mathcal{Z}$ be the set containing all possible values $\mathbf{z}$ of $\mathbf{Z} = \{ Z_{111}, Z_{112}, \ldots, Z_{G I_G n_{G I_G}})^T$. Write $|\mathcal{S}|$ for the number of elements in a finite set $\mathcal{S}$. Then, $|\mathcal{Z}| = \prod_{g=1}^{G} \prod_{i=1}^{I_g} n_{gi}$ since every matched pair $i$ in group $g$ has $n_{gi}$ possible treatment assignment allocations, $(Z_{gi1}, Z_{gi2}, \ldots, Z_{gi n_{gi}}) = (1, 0, \ldots, 0, 0), (0, 1, \ldots, 0, 0)$ or $(0, 0, \ldots, 0, 1)$. In a randomized experiment, a treatment assignment $\mathbf{Z}$ is randomly chosen from $\mathcal{Z}$. Therefore, $\Pr(\mathbf{Z} = \mathbf{z} | \mathcal{F}, \mathcal{Z}) = |\mathcal{Z}|^{-1}$ and $\Pr(\mathbf{Z}_{gij}=1 | \mathcal{F}, \mathcal{Z}) = 1/n_{gi}$ from the independence between strata. The response $\mathbf{R} = (R_{111}, R_{112}, \ldots, R_{G I_G n_{G I_G}})$ is thus random due to $\mathbf{Z}$ whereas $\mathcal{F}$ is fixed. This randomization enables researchers make inference for treatment effects in a randomized experiment (Rosenbaum 2017).

\subsection{Matching and Observational Studies}
\label{ss:observational_studies}

In an observational study, when we collect data, treated and control individuals are not matched. The strata should be formed on the basis of their treatment and covariates. Matching methods are important tools to create the strata. Each treated individual is matched to a control with the same covariates and the same probability of receiving a treatment. In this paper, we consider a matched pair design containing only one control for each treated with $n_{gi}=2$. However, our method that will be discussed throughout this paper can be readily extended to other designs such as matching with multiple controls. In practice, it is difficult to find a control who has the exactly same covariates, especially for continuous covariates. Instead, we find a control as similar to the targeted treated as possible. Then, we assess how similar matched pairs are by checking the overall covariate balance. The most common diagnostic for checking balance is using the standardized difference, see Rosenbaum (2010). The quality of matched pairs produced by matching methods should be assessed and reported before making causal inference. 

Once obtained matched pairs are accepted by passing diagnostics, we can view the matched pairs as pairs in a stratified randomized experiment under the assumption of no unmeasured confounding. This assumption implies that the probability of receiving a treatment $\pi_{gij}=\Pr(\mathbf{Z}_{gij}=1  | \mathbf{x}_{gij})$ depends only on observed covariates $\mathbf{x}_{gij}$ meaning that if two individuals $gij$ and $g'i'j'$ have the same covariates (i.e., $\mathbf{x}_{gij} = \mathbf{x}_{g'i'j'}$), then $\pi_{gij}=\pi_{g'i'j'}$. This property also implies, for a matched pair, $\Pr(\mathbf{Z}_{gi1}=1  | \mathcal{F}, \mathcal{Z})= \Pr(\mathbf{Z}_{gi2}=1  | \mathcal{F}, \mathcal{Z})= 1/2$ since two individuals in a matched pair share the same observed covariates. In addition to the assumption of no unmeasured confounding, another assumption is required to recover the stratified randomized experiments: common support for $\Pr(\mathbf{Z}_{gij}=1  | \mathbf{x}_{gij})$. The common support means that every treated or control individual must have a positive probability of receiving a treatment (and no treatment), that is, $0< \Pr(\mathbf{Z}_{gij}=1  | \mathbf{x}_{gij}) < 1$. Anyone with the probability 1 of receiving a treatment cannot be compared since there exists no control individual who has the same covariates. 

\subsection{Sensitivity to Unmeasured Biases in Observational Studies}
\label{ss:notataion_sensi}

In an observational study, matching methods can adjust for measured confounders, however, it might be possible that two subjects with the same observed covariate have different probabilities of receiving a treatment due to the existence of unmeasured confounders. In the presence of unmeasured confounders, we consider a sensitivity analysis model proposed by Rosenbaum (2002a). This model restricts treatment assignments within a stratum with a sensitivity parameter $\Gamma$. For individuals $gij$ and $g'i'j'$ with $\mathbf{x}_{gij}=\mathbf{x}_{g'i'j'}$, their odds of treatment assignment may differ by at most a factor of $\Gamma \geq 1$, 
\begin{equation}
\frac{1}{\Gamma} \leq \frac{\pi_{gij}\cdot (1-\pi_{g'i'j'})}{(1-\pi_{gij})\cdot \pi_{g'i'j'}} \leq \Gamma
\label{eqn:sensi_model}.
\end{equation}
When $\Gamma=1$, the model~\eqref{eqn:sensi_model} is equivalent to assuming the no unmeasured confounders. The null hypothesis can be conducted by using randomization-based tests, and the $P$-value can be obtained as a point estimate. However, for $\Gamma > 1$, randomization inferences produce an interval of $P$-values. If the endpoints of the interval are less than a significance level $\alpha$, the null hypothesis can be rejected even in the presence of unmeasured confounders since the worst-case $P$-value is less than $\alpha$. The $P$-value interval becomes wider as $\Gamma$ increases, and at a some point, the $P$-value interval contains $\alpha$. When $\alpha$ is in the interval, some $P$-values can reject the null hypothesis, but others cannot reject. Therefore, such $P$-value interval is uninformative, and the null hypothesis cannot be rejected. In practice, it is enough to find the upper bound of the $P$-value interval in order to conduct hypothesis tests. An approximation of the upper $P$-value bound can be used, see Gastwirth, Krieger and Rosenbaum (2000) for more detailed discussions. Using the approximation, it is easy to find the largest value of $\Gamma$ that cannot alter the conclusion that is obtained when $\Gamma=1$. 

\section{A Combined Exploratory and Confirmatory Method for Discovering Effect Modification}
\label{s:method}

\subsection{The Null Hypothesis and the Confidence Interval Method}

Suppose that outcome is continuous. Let $\tau$ be the population average treatment effect. When there is no effect modification at all, every individual $gij$ has the same constant treatment effect, $r_{Tgij} -r_{Cgij}= \tau$. Therefore, to test whether there is effect modification, we can define the null hypothesis as $H_0: r_{Tgij} -r_{Cgij}= \tau$ for every $gij$. Under the null hypothesis, missing potential outcomes can be imputed from the observed data when $\tau$ is a known and fixed value. Then a randomization test can examine how extreme a test statistic is under the null, and thus produce inference. However, $\tau$ is unknown in practice, and is a nuisance parameter that has to be estimated although it is not of primary interest. Difficulties arise because the observed data is used for both estimating $\tau$ and conducting hypothesis tests for discovering effect modification. Berger and Boos (1994) provides an approach to handle this difficulty by maximizing the $P$-value across a confidence interval for $\tau$. Ding et al. (2016) implements this method for testing the null hypothesis in randomized experiment settings, and calls it the \textit{confidence interval} (CI) method. For a brief review of their implementation, the first step is estimating $\tau$ with a $(1-\gamma)$ level confidence interval for a small $\gamma$, for example, $\gamma=0.001$. Hypothesis tests are then conducted for all possible values of $\tau$  in the confidence interval, and the $P$-value is maximized among all obtained $P$-values. Finally, the maximum $P$-value plus $\gamma$ is reported as the $P$-value from the CI method, or the maximum $P$-value is compared with a significance level $\alpha$. 

Our proposed method slightly modifies the CI method. Instead of computing $P$-values across the confidence interval, our method computes test statistics (i.e., ${D}_{\Gamma \max}$ that will be defined in the next subsection) across the confidence interval, finds the minimum of the test statistics, and compare the minimum with the critical value for a significance level $\alpha$. This modified version is referred as the CI method throughout this paper.   

In the following subsection, we assume that $\tau$ is known first, and describe our method for discovering effect modification in observational studies. 

\subsection{Joint Evaluation of Subgroup Comparisons }
\label{ss:joint_eval}


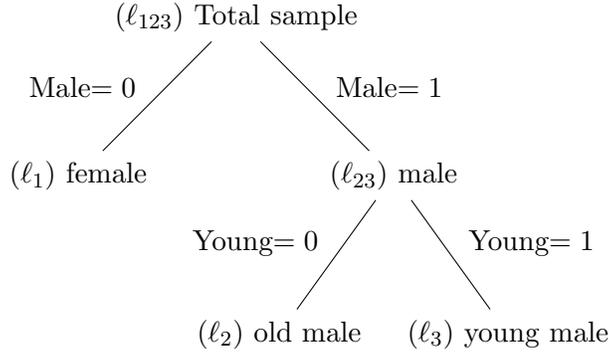
\begin{figure}[t!]
\centering
\begin{tikzpicture}[level distance=60pt, sibling distance=10pt, edge from parent path={(\tikzparentnode) -- (\tikzchildnode)}]
\tikzset{every tree node/.style={align=center}}
\Tree [.{($\ell_{123}$) Total sample} \edge node[auto=right,pos=.6]{Male$=0$}; {($\ell_1$) female} \edge node[auto=left,pos=.6]{Male$=1$};[.{$(\ell_{23}$) male} \edge node[auto=right,pos=.6]{Young$=0$}; {($\ell_2$) old male} \edge node[auto=left,pos=.6]{Young$=1$}; {($\ell_3$) young male}  ]]
\end{tikzpicture}
\caption{An example tree $\Pi$.}
\label{fig:example_tree}
\end{figure}

Suppose that there is a given tree $\Pi$ with $G$ groups that is a partitioning of the covariate space. Each group $g$ represents a terminal node in the partition, and we denote each terminal node as $\ell_g$. To utilize the structure of trees, we trace back how the partition $\Pi$ is built. For each splitting step in building a tree, a certain internal node is chosen and forced to split into two subsequent nodes. This step is repeated until the number of terminal nodes is $G$. Since every step increases the number of nodes by 2 and there are $G-1$ steps, the number of all nodes is $2G-1$. Excluding the initial node, there are $G$ terminal nodes and $G-2$ internal nodes. Each internal node can be constructed as a union of some of $\ell_1, \ldots, \ell_G$. For example, as seen in Figure~\ref{fig:example_tree}, two binary variables, male and young, are used in the tree. The first split is made on the male variable, and the second split is made for male on the young variable. There are three terminal nodes: (1) $\ell_1=$female, (2) $\ell_2=$old male, and (3) $\ell_3=$young male from left to right. The one internal node that represents the male sample can be represented as $\ell_2 \cup \ell_3$. We simply denote this internal node as $\ell_{23}$. Technically, the total sample is one of the internal nodes, but our method will not include this total sample, $\ell_{123}:= \ell_1 \cup \ell_2 \cup \ell_3$, since we want to discover subgroups that have different treatment effects from the population average, and cannot find any effect modification from $\ell_{123}$. In the example, the tree can be represented by $\Pi = \{\{\ell_{23} \}, \{\ell_1\}, \{\ell_2\}, \{\ell_3\} \}$.

With $G$ terminal nodes, $G-2$ internal nodes are considered for {\it de novo} discovering effect modification. It may seem counter-intuitive since considering more comparisons implies paying more for multiple testing. However, the inclusion of internal nodes has several beneficial aspects. We illustrate two beneficial aspects. First, some of terminal nodes may have a small number of matched pairs and consequently lack power for detecting effect modification. Combining some terminal nodes can increase power even though the number of comparisons increase. Second, when a given tree structure is deeper than the true structure, considering only terminal nodes is misleading. This is important especially when $\Pi$ is not given, and has to be estimated. Overfitting a tree leads to a unnecessarily complex structure, but including internal nodes can correct this problem. 

We construct the comparison vector of $(2G-2)$ test statistics for $G$ terminal nodes and $G-2$ internal nodes. Each comparison statistic corresponds to an element in a tree $\Pi$. We can consider the $(2G-2) \times G$ conversion matrix $\mathbf{C}$ that can create the $(2G-2)$ correlated comparisons based on $G$ mutually independent test statistics for $G$ terminal nodes, see Lee et al. (2017) for a discussion of the conversion matrix in a factorial design. To illustrate the matrix $\mathbf{C}$, let us revisit the example shown in Figure~\ref{fig:example_tree}. Including the internal node, there are four nodes. The conversion matrix $\mathbf{C}$ can be constructed as
$$
\mathbf{C}=  \begin{bmatrix}
0 & 1 & 1 \\
1 & 0 & 0 \\
0 & 1 & 0 \\
0 & 0 & 1 
\end{bmatrix}.
$$ 
The first row represents the internal node indicating the male subgroup $\ell_{23}$ and the last three represent the terminal nodes $\ell_1, \ell_2$ and $\ell_3$. Now, let $\mathbf{T} = (T_1, \ldots, T_G)^{T}$ be the vector of the test statistics for terminal nodes. Then, the $(2G-2)$ test statistics for all nodes $\mathbf{S}=(S_1, \ldots, S_{2G-2})^T$ can be obtained as $\mathbf{S}=\mathbf{C}\mathbf{T}$. In the above example, $\mathbf{S} = (T_2+T_3, T_1, T_2, T_3)$, and $T_2+T_3$ represents the test statistic for the male subgroup $\ell_{23}$. 

We consider test statistics of the form $T = \sum_ {g=1}^{G}T_g$ where $T_g$ $=$ $\sum_{i=1}^{I_g}$ $\sum_{j=1}^{2} Z_{gij} q_{gij}$ for suitable scores $q_{gij}$ that are a function of the response $R_{gij}$. Under the null $H_0: r_{Tgij}- r_{Cgij}=\tau$, $R_{gij}$ and $q_{gij}$ are fixed by conditioning on $\mathcal{F}$. The most common statistic of this form is Wilcoxon's signed rank statistic. To test $H_0$, we consider a level $\alpha$ two-sided test since the subgroup treatment effects can be  either larger or smaller than $\tau$. For $\Gamma=1$, randomization inference gives the exact null distribution $\Pr(T_g|\mathcal{F}, \mathcal{Z})$ of the test statistic $T_g$. However, for $\Gamma > 1$, the distribution of $T_g$ is bounded by the distributions of $T_g^{+}$ and $T_g^{-}$ where $E(T_g^{+})= \mu_{\Gamma g}^{+}, E(T_g^{-})= \mu_{\Gamma g}^{-}, \var(T_g^{+})= \nu_{\Gamma g}^{+}$, and $\var(T_g^{-})= \nu_{\Gamma g}^{-}$; see Rosenbaum (2002a). If the treatment effect in subgroup $g$ is larger than $\tau$, $T_g^+$ is used for obtaining the upper bound on the $P$-value for $T_g$; otherwise, $T_g^{-}$ is used. 

For simplicity, we first introduce a one-sided procedure with $T_g^+$ for testing effect modification against a larger effect than $\tau$. A large sample approximation can be applied to the joint distribution of $\mathbf{T}$. Under $H_0$ with mild conditions on $q_{gij}$, the joint distribution of $(T_g-\mu_{\Gamma g}^+)/\sqrt{\nu_{\Gamma g}^+}$, $g=1, \ldots, G$, converges to a multivariate Normal distribution $\mathcal{N}_G(0, \mathbf{I})$ where $\mathbf{I}$ is the $G\times G$ identity matrix as $\min(I_g) \to \infty$. The upper bound on the $P$-value for $T$ can be obtained as $1- \Phi\left\{\left( T - \sum_{g=1}^{G} \mu_{\Gamma g}^+ \right) / \sqrt{\sum_{g=1}^{G} \nu_{\Gamma g}^+} \right\}$. Let $\boldsymbol{\mu}_{\Gamma}^+ = (\mu_{\Gamma 1}^+, \ldots, \mu_{\Gamma G}^+)^T$ and $\mathbf{V}_{\Gamma}^+$ for the $G \times G$ diagonal matrix with $g$-th diagonal element $\nu_{\Gamma g}^+$. Define $\boldsymbol{\theta}_{\Gamma}^+=\mathbf{C}\boldsymbol{\mu}_{\Gamma}^+$ and $\boldsymbol{\Sigma}_{\Gamma}^+=\mathbf{C} \mathbf{V}^+_{\Gamma}\mathbf{C}^{T}$, noting that $\boldsymbol{\Sigma}_{\Gamma}^+$ is not typically diagonal. Write $\theta_{\Gamma k}^+$ for the $k$-th coordinate of $\boldsymbol{\theta}_{\Gamma}$ and $(\sigma_{\Gamma k}^+)^{2}$ for the $k$-th diagonal element of $\boldsymbol{\Sigma}_{\Gamma}^+$. Define $D_{\Gamma k}^+=\left(  S_{k}- \theta_{\Gamma k}^+\right)  /\sigma_{\Gamma k}^+$ and $\mathbf{D}_{\Gamma}^+=\left(  D_{\Gamma 1}^+,\ldots,D_{\Gamma, 2G-2}^+\right)^{T}$. Finally, write $\boldsymbol{\rho}_{\Gamma}^+$ for the $(2G-2)\times (2G-2)$ correlation matrix formed by dividing the element of $\boldsymbol{\Sigma}_{\Gamma}^+$ in row $k$ and column $k^{\prime}$ by $\sigma_{\Gamma k}^+\,\sigma_{\Gamma k^{\prime}}^+$. Under $H_0$, the distribution of $\mathbf{D}_{\Gamma}^+$ converges to a Normal distribution, $\mathcal{N}_{2G-2}\left( \mathbf{0},\boldsymbol{\rho}_{\Gamma}^+ \right)  $, with expectation $\mathbf{0}$ and covariance matrix $\boldsymbol{\rho}_{\Gamma}^+$ as $\min\left( I_{g}\right)  \rightarrow\infty$. Then, the one-sided test can be conducted by using 
\[
D_{\Gamma \max}^+=\max_{1\leq k\leq 2G-2} D_{\Gamma k}^+=\max_{1\leq k\leq 2G-2}  \frac{S_{k}-\theta_{\Gamma k}^+}{\sigma_{\Gamma k}^+} \text{.}
\]
Given a significance level $\alpha/2$, the critical value $\kappa_{\Gamma, \alpha/2}^+$ for $D_{\Gamma \max}^+$ solves 
$$
1-\alpha/2=\Pr\left(  D_{\Gamma \max}^+ <\kappa_{\Gamma, \alpha/2}^+\right)  =\Pr\left(\frac{S_{k}-\theta_{\Gamma k}^+}{\sigma_{\Gamma k}^+}  <\kappa_{\Gamma, \alpha/2}^+,\,k=1,\ldots,2G-2\right) $$
under $H_{0}$. The multivariate Normal approximation to $\kappa_{\Gamma, \alpha/2}^+$ is obtained using the \texttt{qmvnorm} function in the \texttt{mvtnorm} package in R, as applied to the $\mathcal{N}_{2G-2}\left(  \mathbf{0},\boldsymbol{\rho}_{\Gamma}^+\right)$ distribution, see Genz and Bretz (2009).

Similarly, we can define $D_{\Gamma \max}^-$ as the minimum of the deviates $D_{\Gamma k}^-$ that are defined from ${\mu}_{\Gamma g}^-$ and $\nu_{\Gamma g}^-$, that is, $$
D_{\Gamma \max}^-=\min_{1\leq k\leq 2G-2} D_{\Gamma k}^-  =\min_{1\leq k\leq 2G-2}  \frac{S_{k}-\theta_{\Gamma k}^-}{\sigma_{\Gamma k}^-} \text{.}
$$ 
Then, $D_{\Gamma \max}^-$ can be compared with the critical value $\kappa_{\Gamma, \alpha/2}^-$ that solves $1-\alpha/2= \Pr\left(  D_{\Gamma \max}^- > \kappa_{\Gamma, \alpha/2}^-\right)$. The null hypothesis $H_0$ is rejected at a level $\alpha$ when either $D_{\Gamma k}^+ > \kappa_{\Gamma, \alpha/2}^+$ or $D_{\Gamma k}^- < \kappa_{\Gamma, \alpha/2}^-$. 

For Wilcoxon's signed rank statistic and many other statistics, the variances of $T^+$ and $T^-$ are equal, $\nu_{\Gamma g}^+= \nu_{\Gamma g}^-$. This equality implies  $\boldsymbol{V}_{\Gamma}^+=\boldsymbol{V}_{\Gamma}^-$, $\boldsymbol{\rho}_{\Gamma}^+=\boldsymbol{\rho}_{\Gamma}^-$, and further $\kappa_{\Gamma, \alpha/2}^+ = -\kappa_{\Gamma, \alpha/2}^-$. In this case, the two-sided test can be simply conducted by defining 
$$
D_{\Gamma k} = \begin{cases}
D_{\Gamma k}^+ & \text{if} \quad \vert D_{\Gamma k}^+ \vert \geq \vert D_{\Gamma k}^- \vert \\
D_{\Gamma k}^- & \text{if} \quad \vert D_{\Gamma k}^+ \vert < \vert D_{\Gamma k}^- \vert \\
\end{cases}, \quad D_{\Gamma \max} = \max_{1\leq k \leq 2G-2} \vert D_{\Gamma k} \vert.$$
 Then, $D_{\Gamma \max}$ is compared with the common critical value $\kappa_{\Gamma, \alpha} = \vert\kappa_{\Gamma, \alpha/2}^+ \vert = \vert \kappa_{\Gamma, \alpha/2}^- \vert$. For simplicity, this combined two-sided test is considered throughout this paper.

\subsection{Honest Splitting and Existing Methods}

In Section 3.1 and 3.2, we assumed that a tree $\Pi$ is given for discovering effect modification. Trees can be obtained from previous literature, but it is not generally available in many studies. Alternatively, adaptive estimation methods can be used, but the same data is used for building a tree and conducting hypothesis tests. Athey and Imbens (2016) shows that adaptive methods do not have the correct coverage of confidence intervals in the context of estimating treatment effects. They further propose an ``honest'' method, which separates the data into two parts: building a tree and estimating treatment effects. The first sample will not be used for making inference, but used for discovering effect modification structures via recursive partitioning. This separate sample is used for selecting a model structure, thus this method does not assume sparsity. The second sample is used for estimating treatment effects from the discovered structure. 

Compared to Athey and Imben's (2016) method, our method considers the same sample splitting method, but the second sample will be used for conducting hypothesis tests instead of estimation. The discovered tree obtained from the first sample is an exploratory de novo discovery even though it is selected by cross-validation within the first sample. Therefore, if one is interested in inferences of which covariates are effect modifiers other than predictions, a confirmatory method should be accompanied. Joint evaluation with hypothesis testing can reveal the hidden structure in the firstly discovered partitions. Splitting leads to a loss of power for detecting effect modification, but there is a significant benefit for selecting partitions to investigate that offsets the loss. Without splitting, Lee et al. (2017) proposed a factorial design of partitions and a hypothesis testing method under the Fisher's sharp null hypothesis $H_0^{\text{Fisher}}: r_{Tgih} = r_{Cgij}$, not $H_0$. However, as the number of considered covariates increases, statistical power is gradually reduced. Our sample splitting method can be an alternative approach. Since it includes covariate selection, the loss of power of test can be minimized in a high-dimensional setting.

\subsection{Testing the Null Hypothesis for a Subgroup}
\label{ss:sub_hypothesis}

Our primary interest is to test the null hypothesis $H_0: r_{Tgij}-r_{Cgij} = \tau$ for all $g \in \{1, \ldots, G\}$ where $\tau$ is the population mean. The null $H_0$ is a test for effect modification in the whole population. However, testing the null hypothesis $H_0^{{sub}}: r_{Tgij}-r_{Cgij} = \tau$ for a subgroup may be of interest. Rejecting $H_0^{{sub}}$ implies that the corresponding subgroup has a treatment effect significantly different from the population mean. As the test statistic $D_{\Gamma \max}$ is used for testing $H_0$, to test $H_0^{{sub}}$, we consider a test statistic $D_{\Gamma \max}^{sub}$ with respect to the subtree $\Pi^{{sub}}$ that is a subset of $\Pi$ and contains all subsets of the targeted subgroup as elements. Let $\mathcal{I}$ be an index set that indicates the inclusion of $\Pi^{sub}$ in $\Pi$. Since $\vert \Pi \vert = 2G-2$, the index set $\mathcal{I}$ is a subset of $\{1, \ldots, 2G-2 \}$. The new test statistic $D_{\Gamma \max}^{sub} = \max_{k \in \mathcal{I}} D_{\Gamma k}$ can be defined by only focusing on the deviates $D_{\Gamma k}$ for $k \in \mathcal{I}$. Since the number of considered deviates is reduced, it is required to compute a new critical value $\kappa_{\Gamma, \alpha}^{sub}$. The computation can be done by using the sub-correlation matrix $\boldsymbol{\rho}^{sub}_{\Gamma}$ that contains the $(k, k')$ element of $\boldsymbol{\rho}_{\Gamma}$ for $k ,k' \in \mathcal{I}$ that is the intersection of row $k$ and column $k'$. Especially, when $H_0^{sub}$ is for single subgroup $g$, the critical value $\kappa_{\Gamma, \alpha}^{sub}$ can be easily computed as $\Phi^{-1}(1-\alpha/2)$ where $\Phi(\cdot)$ is the cumulative distribution function of the standard Normal distribution. Then, the test statistic $D_{\Gamma \max}^{sub}$ can be compared to $\kappa_{\Gamma, \alpha}^{sub}$. 

\subsection{Parameter Selection in Sensitivity Analysis for Effect Modification}
\label{ss:sensi_analysis}

To assess effect modification in the presence of unmeasured confounding, a sensitivity analysis can be conducted for various values of $\Gamma$. When an unmeasured bias is present, the distribution of $\mathbf{Z}$ is governed by $\Gamma$. The change in the distribution of $\mathbf{Z}$ affects both (a) estimating $100(1-\gamma)$\% the confidence interval and (b) testing the null $H_0$ at a level $\alpha$. For $\Gamma > 1$, as $\Gamma$ increases, the $100(1-\gamma)$\% confidence interval for $\tau$ rapidly converges to the real line even for a comparatively large $\gamma$. When choosing a large value of $\gamma$, the obtained confidence interval may be narrow enough upto a certain $\Gamma$. However, there is a trade-off between $\gamma$ and $\alpha$; a large $\gamma$ means a small $\alpha$ for testing, which may lead to a loss of power. It is difficult to find the optimal balance between $\gamma$ and $\alpha$ since the optimal balance depends on the size of $\Gamma$ that is unknown. To apply the CI method more transparently, we propose to consider $\gamma=0$, which means considering all values of $\tau$ in the real line. For many test statistics of the form $T= \sum_{g=1}^{G} \sum_{i=1}^{I_g} \sum_{j=1}^{2} Z_{gij} q_{gij}$ such as Wilcoxon's signed rank sum test, $D_{\Gamma \max}$ is substantially large when $\tau$ is too small or too large, and the minimum of $D_{\Gamma \max}$ is obtained within a sizable range. In practice, a wide enough range of $\tau$ can be chosen by making sure that $D_{\Gamma \max}$ is large enough at the ends of the range even for a large $\Gamma$. This approach requires more intensive computation, but we may expect a minimal power loss because of $\gamma=0$. 

One may raise the question, ``Can we use $\gamma=0$ all the time in order to maximize the power?'' The answer to this question can be yes or no. If we are only interested in finding evidence of effect modification without further investigation of subgroups, we may use $\gamma=0$. However, $\gamma=0$ should not be used in the subgroup analysis. The real line as the confidence interval for $\tau$ may be too long to provide meaningful inference for testing $H_0^{sub}$ for a certain subgroup. For instance, when $H_0^{sub}$ is for terminal node subgroup $g$, subgroup $g$ is likely not to have effect modification within the subgroup, and thus $H_0^{sub}$ is highly likely to be rejected when $\tau$ is chosen from the real line. Therefore, meaningful causal inference for the subgroup cannot be made when $\gamma=0$. 

\subsection{Binary Outcome}
\label{ss:binary}

When outcomes are binary, the individual treatment effect is $\delta_{gij} = r_{Tgij} - r_{Cgij}$, and $\delta_{gij}$ is an element of $\{-1, 0, 1\}$. The average treatment effect is the average difference between two potential outcomes, denoted by $\delta = (1/N) \sum_{g=1}^{G}\sum_{i=1}^{I_g}\sum_{j=1}^{n_{gi}} \delta_{gij}$ where $N = \sum_{g=1}^{G}\sum_{i=1}^{I_g} n_{gi}$, see Rosenbaum (2001) for further discussion on binary responses. The unbiased estimator of $\delta$ is $\hat{\delta} := \sum_{g=1}^{G} \sum_{i=1}^{I_g} (n_{gi}/N) \hat{\delta}_{gi}$ where $\hat{\delta}_{gi}$ $=$ $\sum_{j=1}^{n_{gi}}$ $\left(Z_{gij}R_{gij}/m_{gij} - (1-Z_{gij})R_{gij}/(n_{gi}-m_{gi})\right)$ is the estimated average treatment effect within stratum $gi$. Also, instead of testing $H_0: r_{Tgij} - r_{Cgij} = \tau$ that is defined for continuous outcomes, we consider a test of the null hypothesis $H_0^{\text{binary}}: \delta = \delta_0$ where $\delta_0 \in \{ d/N: d \in [-N, N] \cap \mathbb{Z} \}$. Since $\delta_{gij}$ can be either -1, 0, or 1, the considered $\delta_0$ has to be a member of $\{ d/N: d \in [-N, N] \cap \mathbb{Z} \}$. Let $\boldsymbol{\delta}=(\delta_{111}, \delta_{112}, \ldots, \delta_{G I_G n_{G I_G}})$. The null $H_0^{\text{binary}}$ allows to test a set $D_{\delta_0} = \{\boldsymbol{\delta} :  \sum_{g=1}^{G}\sum_{i=1}^{I_g}\sum_{j=1}^{n_{gi}} \delta_{gij} =\delta_0 \}$, which means that rejecting $H_0^{\text{binary}}$ is rejecting all $\boldsymbol{\delta}$ in $D_{\delta_0}$. 
 
We adopt Fogarty et al. (2016)'s testing method for binary outcomes, and combine it with our method for discovering effect modification. Let $N_g=\sum_{i=1}^{I_g} n_{gi}$, $\delta_g = (1/N_g)\sum_{i=1}^{I_g}\sum_{j=1}^{n_{gi}} \delta_{gij} $, $\hat{\delta}_g=\sum_{i=1}^{I_g} (n_{gi}/N) \hat{\delta}_{gi}$, and $\Sigma_g = \sum_{i=1}^{I_g} \sigma_{gi}^2$. Now, we define the test statistic vector $\textbf{T}=(T_1, \ldots, T_g)$ where $T_g = N_g \hat{\delta}_g$. A large sample approximation gives that $(T_g - N_g \delta_g)/\sqrt{\Sigma_g}$ has an approximately Normal distribution $N(0, 1)$ for all $g=1, \ldots, G$. For $\Gamma=1$, Fogarty et al. (2016) proposes a method based on randomization inference. To account for worst-case biases, it uses integer programming for finding the maximal variance of $\Sigma_g = \sum_{i=1}^{I_g} \sigma_{gi}^2$ where $\sigma_{gi}^2$ is the variance contribution from stratum $gi$ to $\var(T_g)$. See Section 5 and Theorem 1 in Fogarty et al. (2016) for more details. For $\Gamma > 1$, to find the maximal variance of $\Sigma_g$, a similar approach can be used, but it requires more complicated computations in solving an integer quadratic program, see Fogarty et al. (2017) for detailed computation. The rest of our proposed procedure is the same as the method in Section 3.2 and 3.3. Tree can be discovered based on CART by regressing $\delta_{gij}$ on covariates $\mathbf{x}_{gij}$ from the first sample obtained from sample splitting. From the second sample, the confidence interval for $\delta$ can be constructed by inverting hypothesis tests. Then, the CI method can be applied for each $\delta_0$ in the confidence interval. 

In applying the CI method, however, difficulties arise due to the discreteness of $\delta_0$. When a tree is considered, each terminal node has a different sample size, which causes incompatible hypothesis tests. To illustrate this, consider a simple example with 10 matched pairs ($N=20$) using the tree in Figure~\ref{fig:example_tree}. Suppose that the female subgroup has 5 matched pairs ($N_{\text{female}}=10$). The null for the entire sample is testing whether $\delta_{\text{entire}} = \delta_0$ where $\delta_0$ must be one of $\{ -20/20, -19/20, \ldots, 19/20, 20/20\}$. To discover effect modification, we ultimately want to test $H_0: \delta_{\text{entire}} = \delta_{\text{female}} (= \delta_{\text{old male}} = \delta_{\text{young male}})$. This implies that a test of $\delta_{\text{female}}=\delta_0$ should be considered, however, this test is incompatible with, for instance, $\delta_0 = 3/20$ because $\delta_{\text{female}}$ can be tested only for values $(-10/10, -9/10, \ldots, 10/10)$. A remedy to fix the problem is using two closest compatible values around an incompatible value, conducting hypothesis tests for these two values, and taking a larger $P$-value. For $\delta_0=3/20$, when testing the female group, the two closet compatible values are $1/10$ and $2/10$. This fix is slightly conservative, however, as the number of matched pairs $I_g$ increases, the grid of compatible $\delta_0$ is finer, and the obtained $P$-value converges to the true value. Technically, for each $\delta_0$, let ${\delta}_{g}^{L}$ and $\delta_{g}^{H}$ be the closest two compatible values for subgroup $g$.

\begin{theorem}
\normalfont Under $H_0^{\text{binary}}: \delta = \delta_0$, if $\Sigma_g \to \infty$ as $I_g \to \infty$, then $(N_g \hat{\delta}_g - N_g \delta_g^L)/\sqrt{\Sigma}_g \xrightarrow{d} \mathcal{N}(0,1)$ and $(N_g \hat{\delta}_g - N_g \delta_g^R)/\sqrt{\Sigma}_g \xrightarrow{d} \mathcal{N}(0,1)$. 
\end{theorem}

\noindent \textit{Proof}. As we discussed above, $(N_g \hat{\delta}_g - N_g \delta_0)/\sqrt{\Sigma_g} \xrightarrow{d} \mathcal{N}(0, 1)$ under $H_0: \delta=\delta_0$. Since $I_g \to \infty$, $\left(\delta_g^{R} - \delta_g^{L} \right)=1/N_g$ converges to 0 in probability. Thus, both $\delta_g^{L}$ and $\delta_g^{R}$ converge to $\delta_0$ in probability since ${\delta}_{g}^{L} \leq \delta_0 \leq \delta_g^H$. By Slutsky's theorem, we have that $(N_g \hat{\delta}_g - N_g \delta_g^L)/\sqrt{\Sigma}_g \xrightarrow{d} \mathcal{N}(0,1)$ and $(N_g \hat{\delta}_g - N_g \delta_g^R)/\sqrt{\Sigma}_g \xrightarrow{d} \mathcal{N}(0,1)$ provided that $\Sigma_g \to \infty$. \hfill $\Box$

\section{Simulation}
\label{s:simulation}

In this section, we evaluate the performance of our method with various settings by using simulations in the absence of unmeasured confounding. We consider three main factors that may affect the performance: (1) choice of tree algorithm, (2) split ratio and (3) the degree of effect modification. First, after splitting a sample, the first subsample is used to discover effect modification based on tree algorithms. Two tree approaches may be applied to this discovery step, CART and Causal tree (CT) approaches. Both of the approaches are designed to discover tree structures, however, they have different criteria for constructing the partition and cross-validation, thus provide different partitions. For discussion of CART, see Breiman et al. (1984) and Zhang and Singer (2010), and for discussion of CT, see Athey and Imbens (2016). Second, the splitting ratio of the first subsample to the second may affect the performance. If we invest too much on the first discovery step, we lose power for testing effect modification. On the other hand, if we invest too little, some important structure may not be discovered resulting in loss of power. We consider three ratios (10\%, 90\%), (25\%, 75\%) and (50\%, 50\%). Finally, we examine the performance according the extent of effect modification. If there is small effect modification of some covariates, then our method may not detect this structure from the first subsample and thus may provide low power to discover any effect modification.

We consider two simulation studies, one with continuous outcomes and the other with binary outcomes. For both studies, we set $N=4000$ with 2000 matched pairs and the true effect size as 0.5 on average. Also, we consider five covariates, $x_1, \ldots, x_5$, and assume that at most two of them are true effect modifiers, say $x_1$ and $x_2$. For continuous outcomes, suppose that an individual in stratum of $x_1 =i, x_2=j$ has a treatment effect from a Normal distribution $\mathcal{N}(\tau_{ij}, 1)$. Define $\boldsymbol{\tau} = (\tau_{00}, \tau_{01}, \tau_{10}, \tau_{11})$.  we consider five situations: (1) $\boldsymbol{\tau} = (0.4, 0.4, 0.6, 0.6)$, (2) $\boldsymbol{\tau} = (0.3, 0.3, 0.7, 0.7)$, (3) $\boldsymbol{\tau} = (0.4, 0.4, 0.5, 0.7)$, (4) $\boldsymbol{\tau} = (0.3, 0.3, 0.6, 0.8)$, and (5) $\boldsymbol{\tau} = (0.2, 0.5, 0.5, 0.8)$. For example, the first situation $\boldsymbol{\tau} = (0.4, 0.4, 0.6, 0.6)$ means that there is small effect modification of $x_1$, not $x_2$. The third situation $\boldsymbol{\tau} = (0.4, 0.4, 0.5, 0.7)$ means that there is small effect modification of both $x_1$ and $x_2$. Wilcoxon's signed rank sum test is used for continuous outcomes. Similarly, for binary outcomes, suppose that an individual treatment effect has a binomial distribution $\mathcal{B}(\delta_{ij})$ in stratum $x_1=i, x_2=j$, and define $\boldsymbol{\delta} = (\delta_{00}, \delta_{01}, \delta_{10}, \delta_{11})$. Also, consider five situations: (1) $\boldsymbol{\delta} = (0.45, 0.45, 0.55, 0.55)$, (2) $\boldsymbol{\delta} = (0.4, 0.4, 0.6, 0.6)$, (3) $\boldsymbol{\delta} = (0.45, 0.45, 0.5, 0.6)$, (4) $\boldsymbol{\delta} = (0.4, 0.4, 0.5, 0.7)$, and (5) $\boldsymbol{\delta} = (0.35, 0.5, 0.5, 0.65)$.

\begin{table}[t!]
\centering
\caption{Simulated power (from 10,000 replications) for hypothesis tests to discover effect modification in subgroup analyses. The upper table is for continuous outcomes and the lower table is for binary outcomes. The true effect size is 0.5 on average for the entire population and the sample size is 4000 with 2000 matched pairs.} 
\resizebox{0.9\columnwidth}{!}{%
\begin{tabular}{lll rr rr rr}
\hline
&&& \multicolumn{6}{c}{Splitting ratio} \\
\hline
&&& \multicolumn{2}{c}{(10\%, 90\%)} & \multicolumn{2}{c}{(25\%, 75\%)} & \multicolumn{2}{c}{(50\%, 50\%)}\\
\hline
& Size of effect modification & Continuous outcomes \\
& $(x_1, x_2)$ & $\boldsymbol{\tau} = (\tau_{00}, \tau_{01}, \tau_{10}, \tau_{11})$ & CT & CART & CT & CART & CT & CART \\
\hline
1 & (Small, No) & $ (0.4, 0.4, 0.6, 0.6)$ & 0.09 & 0.04 & 0.10 & 0.05 & 0.07 & 0.06 \\
2 & (Large, No) & $ (0.3, 0.3, 0.7, 0.7)$ & 0.55 & 0.35 & 0.85 & 0.70 & 0.82 & 0.84 \\
3 & (Small, Small) & $(0.4, 0.4, 0.5, 0.7)$ & 0.11 & 0.05 & 0.14 & 0.07 & 0.09 & 0.07\\
4 & (Large, Small) & $(0.3, 0.3, 0.6, 0.8)$ & 0.54 & 0.36 & 0.85 & 0.70 & 0.83 & 0.84\\
5 & (Moderate, Moderate) & $ (0.2, 0.5, 0.5, 0.8)$ & 0.49 & 0.31 & 0.73 & 0.51 & 0.63 & 0.57\\ 
\hline
\hline
& Size of effect modification & Binary outcomes\\
& $(x_1, x_2)$ & $\boldsymbol{\delta} = (\delta_{00}, \delta_{01}, \delta_{10}, \delta_{11})$ & CT & CART & CT & CART & CT & CART \\
\hline
1 & (Small, No) & $ (0.45, 0.45, 0.55, 0.55)$ & 0.05 & 0.02 & 0.05 & 0.03 & 0.04 & 0.03\\
2 & (Large, No) & $ (0.40, 0.40, 0.60, 0.60)$ & 0.61 & 0.33 & 0.79 & 0.69 & 0.76 & 0.78\\
3 & (Small, Small) & $(0.45, 0.45, 0.50, 0.60)$ & 0.07 & 0.02 & 0.07 & 0.04 & 0.05 & 0.04\\
4 & (Large, Small) & $(0.40, 0.40, 0.50, 0.70)$ & 0.68 & 0.38 & 0.90 & 0.77 & 0.90 & 0.89\\
5 & (Moderate, Moderate) & $ (0.35, 0.50, 0.50, 0.65)$ & 0.53 & 0.25 & 0.65 & 0.47 & 0.54 & 0.50\\
\hline
\hline
\end{tabular}
}
\label{tab:sim_conti}
\end{table}

Table~\ref{tab:sim_conti} describes the simulated power of the situations for both continuous and binary outcomes. The upper part of the table shows the simulated power for continuous outcomes. As we expected, when there is small effect modification as the first and third situations, both CT and CART methods produce low power for all three splitting ratios. However, if there is moderate or large effect modification, both can discover effect modification well. The CT method generally has higher power than the CART method. Also, the CT method performs the best with (25\%, 75\%) ratio, however the CART method has the best performance with (50\%, 50\%) ratio. The CART method finds the best fit tree from the first subsample without recognizing the second subsample. If the size of the first subsample is small, it is highly likely that the CART method produces a conservative tree. On the other hand, the CT method accounts for the size of the second subsample, and exploits more exploratory search for tree structures although it often produces false discovery with a high probability. For example, in the first situation with (50\%, 50\%) ratio, the only true effect modifier $x_1$ is discovered in the CT method with probability 0.54 and in the CART method with probability 0.40. The CT method falsely discovers other covariates with probability 0.17, but the CART method with probability 0.07, see Table~\ref{tab:sim_discovery_rate} and Appendix~\ref{appx:sim} for more details on this discovery rates. Although the CT method has a high false discovery rate, falsely discovered partitions will be tested using the second subsample, and will be trimmed after all. The simulated power for binary outcomes is shown in the lower part of Table~\ref{tab:sim_conti}. As we seen in the upper part, for binary outcomes, the CT method also has better performance than the CART method in general. In the analysis of our study that will be discussed in the next section, we will consider (25\%, 75\%) ratio for sample-splitting since this ratio shows the best compromise (measured by power of test) between discovery and confirmation of effect modification .

\section{Causal Effect of Exposure to PM$_{2.5}$ on 5-year Mortality in the New England}
\label{s:example}

\begin{table}[t!]
\centering
\caption{Summary statistics and covariate balance before and after matching.}
\resizebox{0.9\columnwidth}{!}{%
\begin{tabular}{@{\extracolsep{4pt}}lrrrrr@{}}
\hline\\
& & & & \multicolumn{2}{c}{Standardized } \\
& \multicolumn{3}{c}{Summary Statistics} & \multicolumn{2}{c}{Differences}\\
\cline{2-4}\cline{5-6}\\[0.1cm]
& & Control & Control \\
Covariates & Treated &  (Before) & (After) & Before & After \\[0.1cm]
\hline
\textbf{Individual-level} & & \\
\hspace{0.1cm} Male (\%) & 38.5 & 39.9 & 38.5 & -0.02 & 0.00 \\
\hspace{0.1cm} White (\%) & 92.8 & 96.9 & 92.8 & -0.19 & 0.00 \\
\hspace{0.1cm} Medicaid Eligible (\%) & 10.8 & 9.1 & 10.8 & 0.05 & 0.00 \\
\hspace{0.1cm} Age (Group, 1-5) & 2.6 & 2.6 & 2.6 & 0.02 & 0.00 \\
\hspace{0.1cm} Age (65-107) & 76.3 & 76.1 & 76.3 & 0.02 & 0.00 \\[0.3cm]
\textbf{ZIP code-level} & & \\
\hspace{0.1cm} Temperature & 283.5 & 282.9 & 283.4 & 0.55 & 0.06 \\
\hspace{0.1cm} Humidity & 76.1 & 76.9 & 76.1 & -0.44 & 0.01 \\
\hspace{0.1cm} BMI (\%) & 26.1 & 26.3 & 26.1 & -0.44 & -0.06 \\
\hspace{0.1cm} Smoker Rate (\%) & 49.9 & 52.6 & 49.7 & -0.72 & 0.07 \\
\hspace{0.1cm} Black Population (\%) & 6.2 & 3.2 & 6.0 & 0.33 & 0.03 \\
\hspace{0.1cm} Median Household Income & 56.1 & 53.8 & 56.7 & 0.10 & -0.03 \\
\hspace{0.1cm} Median Value of Housing & 207.5 & 184.8 & 205.9 & 0.20 & 0.01 \\
\hspace{0.1cm} \% Below Poverty Level & 8.3 & 9.1 & 8.3 & -0.09 & 0.01 \\
\hspace{0.1cm} \% Below High School Education & 30.6 & 30.1 & 30.2 & 0.03 & 0.03 \\
\hspace{0.1cm} \% of Owner Occupied Housing & 62.9 & 68.9 & 62.7 & -0.33 & 0.01\\
\hspace{0.1cm} Population Density (log-scale) & -6.9 & -8.1 & -7.0 & 0.89 & 0.06 \\
\hline
\end{tabular}
}
\label{tab:cov_balance}
\end{table}

We consider 1,612,414 beneficiaries that enter in the Medicare cohort on January 1 2002 (reference date). For each enrollee, we calculate his/her exposure to PM$_{2.5}$ during the two years prior the entry into the cohort, so from  January 1, 2000 to December 31, 2001.  The outcome is time to death, which can be ascertained up to the end of the study, December 31, 2006. In addition to the exposure and the outcome, we consider both individual-level covariates and ZIP code-level covariates. All covariates are measured in 2001 before the reference date. Each individual provides age, sex (male or female), race (white or non-white) and Medicaid eligibility (a proxy for low socioeconomic status). ZIP code-level covariates consist of temperature, humidity, body mass index (BMI), percentage of ever smokers, black population, median household income, median value of housing, percentage below the poverty level, percentage less than high school education, percentage of owner-occupied housing units, and population density. Table~\ref{tab:cov_balance} displays summary of the treated and control populations. Before matching, treated subjects are more Medicaid eligible, more often female, and more often non-white.

The two year average of PM$_{2.5}$ is obtained in a continuous scale. We create a binary treatment variable using a cutoff value 12 $\mu$g/$m^3$ based on the national standard. In 2012, United States Environmental Protection Agency (EPA) reviewed the national ambient air quality standards for PM$_{2.5}$, and revised the annual PM$_{2.5}$ standard from 15 $\mu$g/$m^3$ to 12 $\mu$g/$m^3$. In this paper, we estimate  the causal effect of being exposed to levels of PM$_{2.5}$ higher than 12 $\mu$g/$m^3$ versus lower than 12 $\mu$g/$m^3$ on 5-year mortality rate. Among 1,612,414 individuals, there are 584,374 treated (i.e., PM$_{2.5}$ $ >$ 12 $\mu$g/$m^3$) and 1,028,040 control (i.e., PM$_{2.5}$ $\leq$ 12 $\mu$g/$m^3$). We note that the level of PM$_{2.5}$ is estimated at the centroid of a ZIP code. Individuals living in the same ZIP code area share the same value of PM$_{2.5}$, thus the same treatment. We use the previously published methods that validate estimation of PM$_{2.5}$ levels. See Di et al. (2016) for more details of estimation methods of exposure to PM$_{2.5}$.

To adjust for measured confounders and discover effect modification, we use a matching method that produces exact matched pairs on four individual-level covariates, white, male, Medicaid eligibility, and age group. The age group variable has 5-year categories of age (1:65-70, 2:71-75, 3:76-80, 4:81-85, and 5:above 85). To obtain exact pairs, the dataset is stratified into $40=2\times 2 \times 2 \times 5$ strata according to levels of individual-level covariates. For each stratum, the ZIP code-level covariates are matched as closely as possible. Matching can be performed by using the \texttt{Optmatch} R package. We randomly select about 20\% of the treated individuals from the entire dataset for a better covariate balance. This allow us to construct 110,091 matched pairs. Covariate balance is shown in Table~\ref{tab:cov_balance}. Since two matched individuals have the same values for individual-level covariates, the standardized differences of them are zero. The standardized differences of ZIP code-level covariates are located between -0.06 and 0.07, which indicates that there is no systematic difference between treated and control.

\begin{figure}[t!]
\centering
\resizebox{\columnwidth}{!}{%
\begin{tikzpicture}[level distance=100pt, edge from parent path={(\tikzparentnode) -- (\tikzchildnode)}]
\tikzset{every tree node/.style={align=center}}
\tikzset{level 1/.style={sibling distance=-100pt}, level 2/.style={sibling distance=-10pt}, level 3/.style={sibling distance=10pt}, level 4/.style={sibling distance=0pt}}
\Tree [.{$(\ell_{1234567})$ Total sample\\ 1.75\% \\(1.27\%, 2.23\%)}
\edge node[auto=right,pos=.6]{\footnotesize Age group$=1, 2, 3$};[.\node[draw]{$(\ell_{123})$ age:65-80\\ 0.41\% \\(-0.13\%, 0.96\%)}; 
\edge node[auto=right,pos=.6]{\footnotesize Age group$=1, 2$};[.\node[draw]{$(\ell_{12})$ age:65-75\\ 0.04\% \\ (-0.59\%, 0.67\%)}; 
\edge node[auto=right,pos=.6]{\footnotesize White$=1$};\node[draw]{$(\ell_1)$ age:65-75,\\ White \\ -0.03\% \\(-0.69\%, 0.62\%)}; 
\edge node[auto=left,pos=.6]{\footnotesize White$=0$};\node[draw, dashed]{$(\ell_2)$ age:65-75,\\ Non-white \\ 0.80\% \\ (-1.40\%, 2.99\%)};]
\edge node[auto=left,pos=.6]{\footnotesize Age group$=3$};\node[draw, dashed]{$(\ell_{3})$ age:76-80 \\ 1.26\% \\(-0.21\%, 2.32\%)};] 
\edge node[auto=left,pos=.6]{\footnotesize Age group$=4, 5$};[.\node[draw]{$(\ell_{4567})$ age: above 80\\ 5.33\% \\(4.31\%, 6.34\%)}; 
\edge node[auto=right,pos=.6]{\footnotesize Age group$=4$};[.\node[draw]{$(\ell_{456})$ age:81-85\\ 3.31\% \\(1.97\%, 4.66\%)}; 
\edge node[auto=right,pos=.6]{\footnotesize Eligible$=0$};[.\node[draw, dashed]{$(\ell_{45})$ age:81-85,\\{Not Medicaid}\\{eligible}\\ 2.73\% \\(1.31\%, 4.16\%)};
 \edge node[auto=right,pos=.6]{\footnotesize Male$=0$};\node[draw, dashed]{$(\ell_4)$ age:81-85,\\{Not Medicaid}\\{eligible, Female} \\ 2.33\% \\(0.57\%, 4.09\%)};
 \edge node[auto=left,pos=.6]{\footnotesize Male$=1$};\node[draw, dashed]{$(\ell_5)$ age:81-85\\{Not Medicaid}\\{eligible, Male} \\ 3.46\% \\(1.05\%, 5.88\%)}; ] 
 \edge node[auto=left,pos=.6]{\footnotesize Eligible$=1$};\node[draw]{$(\ell_6)$ age:81-85\\{Medicaid eligible}\\ 8.04\% \\(3.93\%, 12.15\%)}; ] 
 \edge node[auto=left,pos=.6]{\footnotesize Age group$=5$};\node[draw]{$(\ell_7)$ age:above 85\\ 7.93\% \\(6.39\%, 9.48\%)}; ]]
\end{tikzpicture}
}
\caption{Discovered tree from the first subsample. Actions are represented on edges. Subgroups whose null hypotheses are rejected at a total significance level $\alpha+\gamma=0.05$ are represented by solid rectangles; otherwise, represented by dashed rectangles. The point estimates with the 95\% confidence intervals for the subgroup treatment effects are computed from the second subsample.}
\label{fig:discovered_tree}
\end{figure}
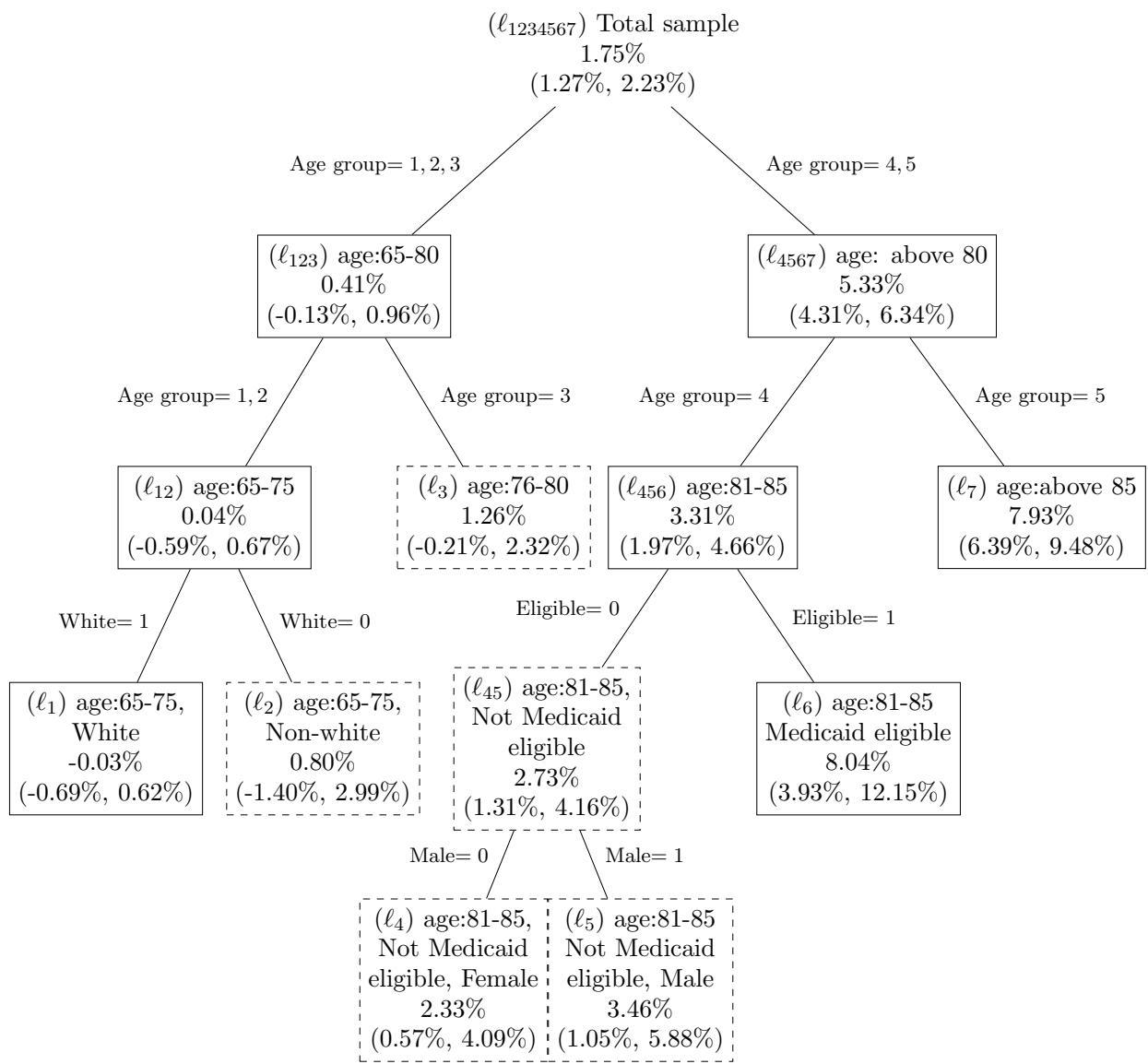

To apply our method, we start with dividing the matched pairs into two subsamples with (25\%, 75\%) ratio. The first subsample of 27,500 matched pairs is used for identifying subgroups with treatment effect heterogeneity, and the other 82,591 matched pairs are used for testing for effect modification. Figure~\ref{fig:discovered_tree} displays the discovered tree structures with seven disjoint subgroups $(\ell_1, ..., \ell_7)$ and five combined subgroups $(\ell_{12}, \ell_{123}, \ell_{45}, \ell_{456}, \ell_{4567})$ from the first subsample (noting that, for example, $\ell_{12}$ is the union of $\ell_1$ and $\ell_2$). The tree $\Pi$ can be represented by $\Pi = \{\{\ell_1\}, \ldots, \{\ell_7\}, \{\ell_{12}\}, \{\ell_{123}\}, \{\ell_{45}\}, \{\ell_{456}\}, \{\ell_{4567}\} \}$ with $\vert \Pi \vert =12$. The first and second splits are made on age group, which creates four age partitions $(\ell_{12}, \ell_{3}, \ell_{456}, \ell_{7})$. The youngest age partition $\ell_{12}$ is further divided by white, and the age partition $\ell_{456}$ is divided by Medicaid eligibility and male. These partitions are obtained from the CT method. The CART method produces a coarser tree with the terminal nodes $(\ell_{12}, \ell_{3}, \ell_{456}, \ell_{7})$. As we found through simulation studies in Section~\ref{s:simulation}, the CART method is slightly more conservative to create subgroups than the CT method. Our general suggestion is that researchers should try both of the methods, and choose a larger tree. 

\begin{table}[t!]
\centering
\caption{Sensitivity analysis for testing the Fisher's hypothesis of no effect: Upper bounds on $P$-values for various $\Gamma$}
\begin{tabular}{c|ccccccc|c}
\hline
& \multicolumn{7}{c|}{Subgroups} & Truncated \\
$\Gamma$ & $\ell_1$ & $\ell_2$ & $\ell_3$ & $\ell_4$ & $\ell_5$ & $\ell_6$ & $\ell_7$ & Product \\
\hline
1.00 & 0.558 & 0.183 & 0.003 & 0.002 & 0.001 & 0.000 & 0.000 & 0.000 \\
1.10 & 1.000 & 0.697 & 0.879 & 0.406 & 0.150 & 0.001 & 0.000 & 0.000 \\
1.20 & 1.000 & 0.965 & 1.000 & 0.987 & 0.812 & 0.024 & 0.000 & 0.000 \\
1.25 & 1.000 & 0.992 & 1.000 & 1.000 & 0.963 & 0.073 & 0.000 & 0.012 \\
1.27 & 1.000 & 0.996 & 1.000 & 1.000 & 0.984 & 0.105 & 0.001 & 0.041 \\
1.28 & 1.000 & 0.997 & 1.000 & 1.000 & 0.989 & 0.124 & 0.001 & 0.068 \\
\hline
\end{tabular}
\label{tab:fisher_sensi}
\end{table}

Before conducting tests for effect modification, we estimate illustrate the treatment effect in the entire population. The obtained matched pairs can be used for testing the Fisher's hypothesis of no effect $H_0^{\text{Fisher}}: r_{Tgij}=r_{Cgij}$. The rejection of $H_0^{\text{Fisher}}$ implies that the exposure to high-level PM$_{2.5}$ is significantly detrimental to health. We consider the truncated product method proposed by Hsu et al.(2013) with the seven discovered subgroups $(\ell_1, \ldots, \ell_7)$ in Figure~\ref{fig:discovered_tree}. This method computes upper bounds on $P$-values for each of the seven subgroups, and then combines the $P$-values using the truncated product proposed by Zaykin et al. (2002). The null hypothesis $H_0^{\text{Fisher}}$ can be tested by using McNemar tests with the second subsample of 82,591 matched pairs. Table~\ref{tab:fisher_sensi} shows the sensitivity analysis with Upper bounds on $P$-values for the Medicare data. At $\Gamma=1$, $H_0^{\text{Fisher}}$ is rejected with a one-sided $P$-value of 4.1$\times 10^{-20}$. At $\Gamma=1.1$, $H_0^{\text{Fisher}}$ is rejected at the $2.2 \times 10^{-13}$, and at $\Gamma=1.2$, $H_0^{\text{Fisher}}$ is still rejected at the $1.0 \times 10^{-4}$. Also, at $\Gamma =1.27$, the Fisher's hypothesis of no effect is rejected at the 0.041 level, but at $\Gamma=1.28$, the hypothesis is not rejected at the 0.05 level. Therefore, we can conclude that exposure to high-level PM$_{2.5}$ increased the 5-year mortality rate even in the presence of unmeasured biases up to $\Gamma=1.27$. 
 Furthermore, the sensitivity parameter $\Gamma$ can be represented as a curve of two parameters $(\Lambda, \Delta)$. Technically, $\Gamma = (\Delta \Lambda +1)/(\Delta + \Lambda)$, see Rosenbaum and Silber (2009). The parameter $\Lambda$ describes the relationship between an unmeasured confounder $u_{gij}$ and treatment assignment $Z_{gij}$, and the parameter $\Delta$ describes the relationship between $u_{gij}$ and the potential outcome $r_{Cgij}$. For example, $\Gamma=1.27$ corresponds to $\Lambda=2.11$ and $\Delta =2$. To illustrate this, consider an unmeasured variable $u_{gij}$ of time spent outdoors that is negatively associated with both the treatment and the outcome. Here, $(\Lambda, \Delta)=(2.11, 2)$ implies that $u_{gij}$ doubles the odds of exposure to high-level PM$_{2.5}$ and increases the odds of death by 2.11-fold. Our sensitivity analysis claims that the conclusion remains even in the presence of any $u_{gij}$ with $(\Lambda, \Delta)$ satisfying $(\Delta \Lambda +1)/(\Delta + \Lambda) \leq 1.27$.

\begin{table}[t!]
\centering
\caption{Sensitivity analysis for testing the null hypothesis of no effect modification and descriptions of the discovered subgroups. The upper table shows twelve deviates from the subgroups with the maximum absolute deviate where the critical values $\kappa_{\Gamma, \alpha}=2.85$ for $\Gamma=1$ when $\alpha=0.04$ and $\gamma=0.01$ and $\kappa_{\Gamma, \alpha}=2.78$ for $\Gamma>1$ when $\alpha=0.05$ and $\gamma=0$, and the lower table shows the proportions of the subgroups and comparisons of outcomes between treated and control.}
\resizebox{0.95\columnwidth}{!}{%
\begin{tabular}{lc|rrrrrrr rrrrr|r }
\hline
& & \multicolumn{12}{c|}{Subgroups} & \\
&  & $\ell_1$ & $\ell_2$ & $\ell_3$ & $\ell_4$ & $\ell_5$ & $\ell_6$ & $\ell_7$ & $\ell_{12}$ & $\ell_{123}$ & $\ell_{45}$ & $\ell_{456}$ & $\ell_{4567}$ &  \\
\hline
$\Gamma$ &$\delta_0$ & $D_{\Gamma 1}$ & $D_{\Gamma 2}$ & $D_{\Gamma 3}$ & $D_{\Gamma 4}$& $D_{\Gamma 5}$ & $D_{\Gamma 6}$ & $D_{\Gamma 7}$ & $D_{\Gamma 8}$ & $D_{\Gamma 9}$ & $D_{\Gamma 10}$ & $D_{\Gamma 11}$ & $D_{\Gamma 12}$ & $D_{\Gamma \max}$ \\
\hline
1 & 0.0112 &-3.43 &-0.29 & 0.27 & 1.35 &1.90 &3.30 &8.66 &-3.37 &-2.55 & 2.23 & 3.20 & 8.14 & 8.66\\
& 0.0143 &-4.37 &-0.57 &-0.31 & 1.00 &1.66 &3.16 &8.25 &-4.34 &-3.68 & 1.80 & 2.75 & 7.54 & 8.25\\
& 0.0175 &-5.31 &-0.85 &-0.91 & 0.64 &1.40 &2.99 &7.85 &-5.33 &-4.83 & 1.36 & 2.28 & 6.92 & 7.85\\
& 0.0207 &-6.27 &-1.13 &-1.50 & 0.29 &1.13 &2.85 &7.45 &-6.32 &-5.97 & 0.92 & 1.82 & 6.31 & 7.45\\
& 0.0238 &-7.20 &-1.41 &-2.09 &-0.06 &0.87 &2.71 &7.05 &-7.30 &-7.11 & 0.48 & 1.36 & 5.70 & 7.30\\[0.2cm]

1.010 & 0.0242 & -6.39 & -1.16 & -1.52 & 0.00 & 0.55 & 2.50 & 6.53 & -6.45 & -6.10 & 0.44 & 1.46 & 5.85 & 6.53\\
1.020 & 0.0254 & -5.85 & -1.01 & -1.12 & 0.00 & 0.16 & 2.26 & 5.91 & -5.89 & -5.40 & 0.13 & 1.09 & 5.17 & 5.91\\
1.030 & 0.0266 & -5.31 & -0.84 & -0.71 & 0.00 & 0.00 & 2.05 & 5.30 & -5.32 & -4.70 & 0.00 & 0.97 & 4.76 & 5.32\\
1.040 & 0.0275 & -4.68 & -0.65 & -0.27 & 0.00 & 0.00 & 1.82 & 4.73 & -4.67 & -3.91 & 0.00 & 0.93 & 4.36 & 4.73\\
1.050 & 0.0285 & -4.07 & -0.45 &  0.00 & 0.00 & 0.00 & 1.61 & 4.17 & -4.03 & -3.26 & 0.00 & 0.80 & 3.80 & 4.17\\
1.060 & 0.0297 & -3.56 & -0.32 &  0.00 & 0.00 & 0.00 & 1.41 & 3.58 & -3.50 & -2.86 & 0.00 & 0.73 & 3.33 & 3.58\\
1.070 & 0.0306 & -2.97 & -0.13 &  0.00 & 0.00 & 0.00 & 1.18 & 3.02 & -2.87 & -2.37 & 0.00 & 0.63 & 2.84 & 3.02\\
1.074 & 0.0312 & -2.80 & -0.08 &  0.00 & 0.00 & 0.00 & 1.08 & 2.77 & -2.70 & -2.24 & 0.00 & 0.59 & 2.63 & 2.80\\
1.075 & 0.0312 & -2.72 & -0.05 &  0.00 & 0.00 & 0.00 & 1.06 & 2.73 & -2.61 & -2.16 & 0.00 & 0.58 & 2.57 & 2.73\\
1.080 & 0.0315 & -2.38 &  0.00 &  0.00 & 0.00 & 0.00 & 0.98 & 2.47 & -2.28 & -1.90 & 0.00 & 0.54 & 2.35 & 2.47\\
\hline
\hline
& & \multicolumn{12}{c|}{Subgroups} & Total \\
\hline 
\multicolumn{2}{l|}{Proportion (\%)} & 46.2 & 4.2 & 22.3 & 8.8 & 4.9 & 1.7 & 11.9 & 50.4 & 72.8 & 13.7 & 15.4 & 27.2 & 100.0 \\
\multicolumn{2}{l|}{Treated (\%)} & 14.5 & 15.2 & 26.5 & 35.5 & 46.8 & 54.6 & 61.7 & 14.6 & 18.3 & 39.6 & 41.2 & 50.1 & 26.9\\
\multicolumn{2}{l|}{Control (\%)} & 14.6 & 14.4 & 25.3 & 33.2 & 43.4 & 46.6 & 53.7 & 14.6 & 17.8 & 36.8 & 37.9 & 44.8 & 25.2 \\
\multicolumn{2}{l|}{Risk difference (\%)} & 0.0 & 0.8 & 1.3 & 2.3 & 3.5 & 8.0 & 7.9 & 0.0 & 0.4 & 2.7 & 3.3 & 5.3 & 1.8 \\
\multicolumn{2}{l|}{Odds ratio} & 1.00 & 1.07 & 1.07 & 1.11 & 1.15 & 1.38 & 1.39 & 1.00 & 1.03 & 1.12 & 1.15 & 1.24 & 1.10\\
\hline
\end{tabular}
}
\label{tab:deviates}
\end{table}

 Returning to testing the null hypothesis $H_0$ of no effect modification, the second subsample is used for confirming and identifying effect modification in the discovered tree structures shown in Figure~\ref{fig:discovered_tree}. Since we do not know the true value of the population average of $\delta$, we first estimate the $100(1-\gamma)\%$ confidence interval for $\delta$ with $\gamma=0.01$, $(1.12\%, 2.38\%)$ and apply the CI method for testing the global null hypothesis $H_0$ of no effect modification. Table~\ref{tab:deviates} shows twelve deviates from the discovered subgroups for various $\delta_0$ at $\Gamma=1$. A negative deviate means that the corresponding subgroup treatment effect is below the population average, and a positive deviate means the opposite. The critical value $\kappa_{\Gamma, \alpha}$ is almost constant as $\kappa_{\Gamma, \alpha}=2.85$ at $\Gamma=1$, and is obtained from the multivariate Normal distribution with $\alpha=0.04$ to achieve a total significance level $\alpha+\gamma=0.05$. At $\Gamma=1$, the maximum absolute deviate $D_{\Gamma \max}$ varies from 7.30 to 8.66, and is always larger than $\kappa_{\Gamma, \alpha}$. This indicates that there is statistically significant effect modification in the entire population when there is no unmeasured confounding. 

 In addition, one may be interested in testing the null hypothesis $H_0^{sub}$ for a certain subgroup that the treatment effect in the subgroup is the same as the population average treatment effect. For instance, policymakers may want to know whether Medicare beneficiaries aged between 81-85, $\ell_{456}$,  are at a high risk of death. To test this, we can focus on the subset of deviates $\{D_{\Gamma 4}, D_{\Gamma 5}, D_{\Gamma 6}, D_{\Gamma 10}, D_{\Gamma 11}\}$, which means $\mathcal{I}=\{4,5,6,10,11\}$. A new critical value $\kappa_{\Gamma, \alpha}^{sub}$ is 2.56 that is smaller than $\kappa_{\Gamma, \alpha}=2.85$. At $\Gamma=1$, the null hypothesis $H_0^{sub}$ for $\ell_{456}$ is rejected since $D_{\Gamma \max}^{sub} = D_{\Gamma 6}$ and $D_{\Gamma 6}$ exceeds 2.56 for all values of $\delta_0$ in the interval. For the terminal nodes such as $\ell_1, \ldots, \ell_7$, $H_0^{sub}$ can be tested with the critical value $\kappa_{\Gamma, \alpha}^{sub}=2.05$ obtained from the standard Normal distribution with $\alpha=0.04$. Figure~\ref{fig:discovered_tree} represents subgroups with solid rectangles whose null hypotheses are rejected at $\Gamma=1$, otherwise, with dashed rectangles. The subgroup $\ell_1$ (white, aged between 65-75) has the treatment effect size significantly lower than the population average, but the subgroups $\ell_6$ and $\ell_7$ have the effect sizes significantly higher than the population average. Also, in Figure~\ref{fig:discovered_tree}, the point estimates and the 95\% confidence intervals for subgroup treatment effects are displayed. We note that each subgroup's confidence interval is computed by inverting the null hypothesis for the subgroup; for example, the confidence interval for $\ell_1$ is an inversion of testing the null hypothesis $H_0: \delta= \delta_{1}$, not testing $H_0: \delta = \delta_0$, where $\delta_1$ is the average treatment effect within stratum $\ell_1$. The lower part of Table~\ref{tab:deviates} provides the detailed descriptions of the discovered subgroups.

\begin{figure}[t!]
\centering
\includegraphics[width=120mm]{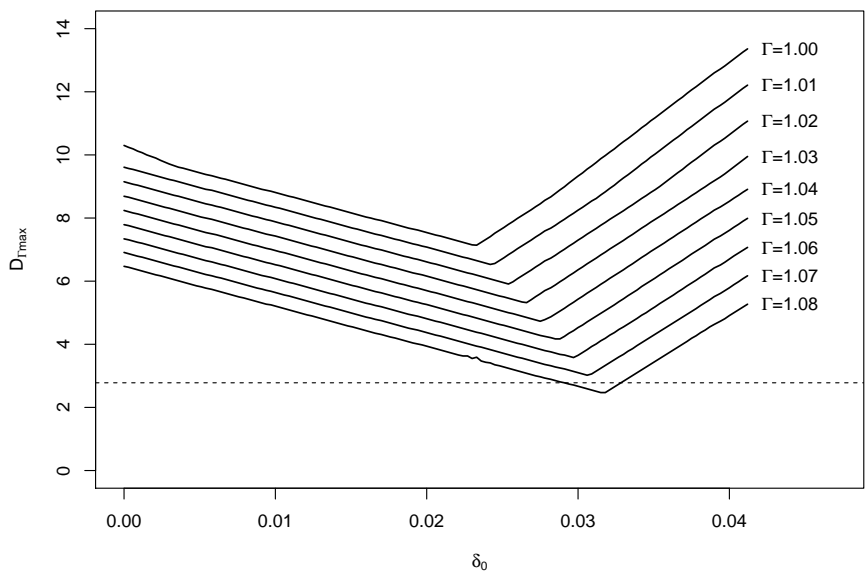}
\caption{The maximum absolute deviate $D_{\Gamma \max}$ for various $\Gamma$ in the interval [0, 0.04] of $\delta$. The dashed line represents the critical value $\kappa_{\Gamma, \alpha}=2.78$}
\label{fig:sensi}
\end{figure}

Table~\ref{tab:deviates} performs a sensitivity analysis for unmeasured confounding in testing effect modification. Since we set $\gamma=0$ for $\Gamma >1$, the critical value $\kappa_{\Gamma, \alpha}=2.78$ is obtained at $\alpha=0.05$. For each value of $\Gamma$, only the minimum of $D_{\Gamma, \max}$ is reported in the table. For example, at $\Gamma = 1.01$, $D_{\Gamma \max}$ can be computed for $\delta_0$ in $(-\infty, \infty)$, and has the minimum 6.53 at $\delta_0 = 0.0242$. $D_{\Gamma \max}$ is attained at either the deviate $D_{\Gamma 1}$ or the deviate $D_{\Gamma 7}$. Therefore, it can be inferred that the subgroups $\ell_1$ and $\ell_7$ have the least sensitivity to unmeasured biases. Figure~\ref{fig:sensi} displays the maximum absolute deviate $D_{\Gamma \max}$ across $\delta_0$ in the interval [0, 0.04] for each value of $\Gamma$. The curve of $D_{\Gamma \max}$ has a V-shape. All the curves have the minimum within the interval, and for $\Gamma \leq 1.07$, the curves are above the horizontal line of the critical value $\kappa_{\Gamma, \alpha}=2.78$. This implies that the null hypothesis $H_0$ of no effect modification is rejected only up to $\Gamma \leq 1.07$. Table~\ref{tab:deviates} shows more calibrated values of $\Gamma$ for a sensitivity analysis. As shown in the table, $D_{\Gamma \max}$ is larger than $\kappa_{\Gamma, \alpha}=2.78$ until $\Gamma=1.074$. This sensitivity analysis shows that there is statistically significant evidence of effect modification if an unmeasured bias does not exceed $\Gamma=1.074$. A bias of $\Gamma=1.074$ corresponds to an unobserved covariate that increases the odds of exposure to high level PM$_{2.5}$ by 1.5-fold and increases the odds of death by more than 1.434-fold (i.e., $(\Delta, \Lambda) = (1.5, 1.434))$.

\section{Discussion}
\label{s:discussion}

Our method discovers effect modification by putting balanced efforts into exploratory and confirmatory discoveries. Instead of determining a set of covariates a priori before making inference, the exploratory search can reveal the structure of effect modification as a form of a tree with some selected subgroups. Then, hypothesis testing based on randomization inference is conducted to confirm whether there is significant evidence of effect modification. We also developed a sensitivity analysis to assess the effect of unmeasured biases on the conclusion, which was not considered in previous studies.
From the Medicare data in the New England, first we found evidence that exposure to PM$_{2.5}$ significantly increases the 5-year morality rate. Sensitivity analysis results showed that the evidence is quite insensitive to unmeasured biases. In addition to making inference about the treatment effect for the entire population, we found evidence that the subgroup treatment effects vary across the population. We discovered that Medicaid eligible seniors between 81-85 and seniors above 85 experienced significantly higher 5-year mortality rates than the population average. Also, we discovered that the group of white and age between 65-75 has a significantly lower mortality rate than the population average. The conclusion remained same if there is no unmeasured bias of $\Gamma>1.074$, which was supported by the sensitivity analysis. Also, it is worth noting that our method can be applied to both continuous and binary outcome settings.

Our method considers the sample-splitting approach that divides the entire sample into two subsamples. However, there has been little literature to select the optimal splitting ratio. Specifically, when applying our method, it is not known what ratio can provide the highest power of test. We considered three ratios through simulation studies in Section~\ref{s:simulation} to decide the optimal ratio among them, and found that the optimal ratio among the tree ratios depending on the size of effect modification. However, the simulation results cannot be a general guideline for those who do not have any prior knowledge from literature about how large effect modification might be. Selecting the optimal ratio without any prior information can be an interesting problem for future research.


\appendix
\section{Discovery Rates of Covariates}
\label{appx:sim}

\begin{table}[t!]
\centering
\caption{Simulated discovery rates of covariates from the first subsample. The upper table is for continuous outcomes and the lower table is for binary outcomes. The sample size is 4000 with 2000 matched pairs.} 
\resizebox{0.9\columnwidth}{!}{%
\begin{tabular}{llcc rr rr rr}
\hline
&&&& \multicolumn{6}{c}{Splitting ratio} \\
\hline
&&&& \multicolumn{2}{c}{(10\%, 90\%)} & \multicolumn{2}{c}{(25\%, 75\%)} & \multicolumn{2}{c}{(50\%, 50\%)}\\
\hline
& Size of effect & \\
Continuous & modification & True effect &  & \multicolumn{6}{c}{Discovery rate}\\
outcomes & $(x_1, x_2)$ & modifier?& Covariate & CT & CART & CT & CART & CT & CART \\
\hline
1 & (S, No) & Y & $x_1$ & 0.29 & 0.13 & 0.53 & 0.21 & 0.54 & 0.40\\
& & N & $x_2$ & 0.16 & 0.07 & 0.29 & 0.07 & 0.17 & 0.07\\
2 & (L, No) & Y & $x_1$ & 0.56 & 0.37 & 0.90 & 0.72 & 0.97 & 0.96\\
& & N & $x_2$ & 0.15 & 0.10 & 0.34 & 0.08 & 0.19 & 0.07\\
3 & (S, S) & Y & $x_1$ & 0.29 & 0.15 & 0.55 & 0.24 & 0.57 & 0.43\\
& & Y & $x_2$ & 0.20 & 0.10 & 0.40 & 0.14 & 0.37 & 0.22\\
& & N & $x_3$ & 0.16 & 0.09 & 0.31 & 0.08 & 0.20 & 0.09\\
4 & (L, S) & Y & $x_1$ & 0.55 & 0.37 & 0.89 & 0.72 & 0.97 & 0.96\\
& & Y & $x_2$ & 0.19 & 0.14 & 0.46 & 0.18 & 0.41 & 0.23\\
& & N & $x_3$ & 0.14 & 0.10 & 0.34 & 0.10 & 0.22 & 0.08\\
5 & (M, M) & Y & $x_1$ & 0.37 & 0.25 & 0.75 & 0.46 & 0.84 & 0.72\\
& & Y & $x_2$ & 0.37 & 0.25 & 0.75 & 0.45 & 0.84 & 0.72\\
& & N & $x_3$ & 0.13 & 0.11 & 0.36 & 0.13 & 0.31 & 0.16\\
\hline
\hline
& Size of effect & \\
Binary & modification & True effect &  & \multicolumn{6}{c}{Discovery rate}\\
outcomes & $(x_1, x_2)$ & modifier?& Covariate & CT & CART & CT & CART & CT & CART \\
\hline
1 & (S, No) & Y & $x_1$ & 0.40 & 0.12 & 0.54 & 0.22 & 0.51 & 0.40\\
& & N & $x_2$ & 0.28 & 0.06 & 0.34 & 0.07 & 0.15 & 0.08\\
2 & (L, No) & Y & $x_1$ & 0.69 & 0.35 & 0.90 & 0.73 & 0.98 & 0.96\\
& & N & $x_2$ & 0.33 & 0.09 & 0.38 & 0.08 & 0.15 & 0.07\\
3 & (S, S) & Y & $x_1$ & 0.42 & 0.13 & 0.56 & 0.23 & 0.56 & 0.44\\
& & Y & $x_2$ & 0.33 & 0.08 & 0.44 & 0.13 & 0.35 & 0.23\\
& & N & $x_3$ & 0.30 & 0.07 & 0.36 & 0.07 & 0.18 & 0.10\\
4 & (L, S) & Y & $x_1$ & 0.72 & 0.39 & 0.94 & 0.79 & 0.99 & 0.98\\
& & Y & $x_2$ & 0.52 & 0.21 & 0.77 & 0.46 & 0.84 & 0.75\\
& & N & $x_3$ & 0.34 & 0.11 & 0.47 & 0.12 & 0.24 & 0.12\\
5 & (M, M) & Y & $x_1$ & 0.55 & 0.22 & 0.79 & 0.46 & 0.84 & 0.74\\
& & Y & $x_2$ & 0.55 & 0.23 & 0.78 & 0.46 & 0.84 & 0.74\\
& & N & $x_3$ & 0.33 & 0.10 & 0.47 & 0.13 & 0.29 & 0.16\\
\hline
\hline
\end{tabular}
}
\label{tab:sim_discovery_rate}
\end{table}

In this section, we discuss the rates of discovering the correct structure of effect modification. Both the CT and CART method are compared in simulation studies. Using the same simulation setting as Section~\ref{s:simulation}, Table~\ref{tab:sim_discovery_rate} reports simulated discovery rates of considered covariates for various splitting ratios with $N=4000$ from 1000 simulated datasets. The rates are obtained by using the first subsample in the sample-splitting approach. For example, for (10\%, 90\%) ratio, a tree is discovered from the first subsample of size 400. The upper table is for continuous outcomes, and the lower table is for binary outcomes. For situations 1 and 2, only $x_1$ is an effect modifier. Table~\ref{tab:sim_discovery_rate} only reports the rates for $x_1$ and $x_2$. The rate for $x_2$ indicates the false discovery rate. Other rates for $x_3$, $x_4$, and $x_5$ are similar, so they are omitted in the table. Similarly, $x_1$ and $x_2$ are effect modifiers for situations 3, 4, and 5. Table~\ref{tab:sim_discovery_rate} shows the rates for $x_1$, $x_2$, and $x_3$, and the rate for $x_3$ represents the false discovery rate. As shown in the table, the CT method produces a more exploratory search than the CART method in every case. Although it can often falsely discover incorrect effect modifiers, it finds the correct effect modifiers with high probabilities. Since the second confirmation subsample is applied to trim excessive findings, this exploratory search results in the increase of power of test as we discussed in Section~\ref{s:simulation}.

\section*{Reference}
\smallskip\setlength{\hangindent}{12pt}
\noindent
Athey, S., and Imbens, G. (2016), ``Recursive Partitioning for Heterogeneous Causal Effects,'' \textit{Proceedings of the National Academy of Sciences}, 113, 7353--7360.

\smallskip\setlength{\hangindent}{12pt}
\noindent
Berger, R. L., and Boos, D. D. (1994), ``P Values Maximized Over a Confidence Set for the Nuisance Parameter,'' \textit{Journal of the American Statistical Association}, 89, 1012--1016. 

\smallskip\setlength{\hangindent}{12pt}
\noindent
Breiman, L., Friedman, J. H., Olshen, R. A., and Stone, C. J. (1984), \textit{Classification and Regression Trees}, California: Wadsworth.

\smallskip\setlength{\hangindent}{12pt}
\noindent
Di, Q., Kloog, I., Koutrakis, P., Lyapustin, A., Wang, Y., and Schwartz, J. (2016), ''Assessing PM$_{2.5}$ Exposures with High Spatiotemporal Resolution across the Continental United States,'' \textit{Environmental Science \& Technology}, 50, 4712--4721.

\smallskip\setlength{\hangindent}{12pt}
\noindent
Di, Q., Wang, Y., Zanobetti, A., Wang, Y., Koutrakis, P., Choirat, C., Dominici, F., and Schwartz, J. (2017), ''Air Pollution and Mortality in the Medicare Population,'' \textit{The New England Journal of Medicine}, 376, 2513--2522. 

\smallskip\setlength{\hangindent}{12pt}
\noindent
Ding, P., Feller, A., and Miratrix, L. (2016), ``Randomization Inference for Treatment Effect Variation,'' \textit{Journal of the Royal Statistical Society: Series B}, 78, 655--671. 

\smallskip\setlength{\hangindent}{12pt}
\noindent
Dockery, D. W., Pope, C. A., Xu, X., Spengler, J. D., Ware, J. H., Fay, M. E., Ferris, B. G., and Speizer, F. E. (1993), ``An Association between Air Pollution and Mortality in Six U.S. Cities,'' \textit{The New England Journal of Medicine}, 329, 1753--1759. 

\smallskip\setlength{\hangindent}{12pt}
\noindent
Dominici, F., Peng, R. D., Bell, M. L., Pham, L., McDermott, A., Zeger, S. L., and Samet, J. M. (2006), ``Fine Particulate Air Pollution and Hospital Admission for Cardiovascular and Respiratory Diseases,'' \textit{JAMA}, 295, 1127--1134. 

\smallskip\setlength{\hangindent}{12pt}
\noindent
Fogarty, C. B., Mikkelsen, M. E., Gaieski, D. F., and Small, D. S. (2016), ``Discrete Optimization for Interpretable Study Populations and Randomization Inference in an Observational Study of Severe Sepsis Mortality,'' \textit{Journal of the American Statistical Association}, 111, 447--458. 

\smallskip\setlength{\hangindent}{12pt}
\noindent
Fogarty, C. B., Shi, P., Mikkelsen, M. E., and Small, D. S. (2017), ``Randomization Inference and Sensitivity Analysis for Composite Null Hypotheses with Binary Outcomes in Matched Observational Studies,'' \textit{Journal of the American Statistical Association}, 112, 321--331. 

\smallskip\setlength{\hangindent}{12pt}
\noindent
Gastwirth, J. L., Krieger, A. M., and Rosenbaum, P. R. (2000), ``Asymptotic Separability in Sensitivity Analysis,'' \textit{Journal of the Royal Statistical Society, Series B,} 62, 545--555.

\smallskip\setlength{\hangindent}{12pt}
\noindent
Genz, A., and Bretz, F. (2009), {\it Computation of Multivariate Normal and t Probabilities}, New York: Springer. (\texttt{R} package \texttt{mvtnorm})

\smallskip\setlength{\hangindent}{12pt}
\noindent
Hansen, B. B. (2004), ``Full Matching in an Observational Study of Coaching for the SAT,'' \textit{Journal of the American Statistical Association}, 99, 609--618. 

\smallskip\setlength{\hangindent}{12pt}
\noindent
Hsu, J. Y., Small, D. S., and Rosenbaum P. R. (2013), ``Effect Modification and Design Sensitivity in Observational Studies,'' \textit{Journal of the American Statistical Association}, 108, 135--148. 


\smallskip\setlength{\hangindent}{12pt}
\noindent
Lee, K., Small, D. S., and Rosenbaum, P. R. (2017), ``A Powerful Approach to the Study of Moderate Effect Modification in Observational Studies,'' \textit{arXiv: 1702.00525}. 

\smallskip\setlength{\hangindent}{12pt}
\noindent
Loomis, D., Grosse, Y, Lauby-Secretan, B., El Ghissassi, F, Bouvard, V., Benbrahim-Tallaa, L., Guha, N., Baan, R., Mattock, H., and Straif, K. (2013), ``The Carcinogenicity of Outdoor Air Pollution,'' \textit{The Lancet Oncology}, 14, 1262--1263. 

\smallskip\setlength{\hangindent}{12pt}
\noindent
Makar, M., Antonelli, J., Di, Q., Cutler, D., Schwartz, J., and Dominici, F. (2017), ``Estimating the Causal Effect of Low Levels of Fine Particulate Matter on Hospitalization,'' \textit{Epidemiology}, 28, 627--634.

\smallskip\setlength{\hangindent}{12pt}
\noindent
Neyman, J. (1923), ``On the Application of Probability Theory to Agricultural Experiments. Essay on Priciples. Section 9 (in Polish),'' \textit{Roczniki Nauk Rolniczych Tom X}, 1--51. Reprinted in \textit{Statistical Science}, 1990, 5, 463--480. 

\smallskip\setlength{\hangindent}{12pt}
\noindent
Rosenbaum, P. R. (2001), ``Effects Attributable to Treatment: Interference in Experiments and Observational Studies With a Discrete Pivot,'' \textit{Biometrika}, 88, 219--231. 

\smallskip\setlength{\hangindent}{12pt}
\noindent
------(2002a), \textit{Observational Studies}, New York: Springer. 

\smallskip\setlength{\hangindent}{12pt}
\noindent
------(2002b), ``Covariance Adjustment in Randomized Experiments and Observational Studies,'' \textit{Statistical Science}, 17, 286--327.

\smallskip\setlength{\hangindent}{12pt}
\noindent
------(2010), \textit{Design of Observational Studies}, New York: Springer. 

\smallskip\setlength{\hangindent}{12pt}
\noindent
------(2017), \textit{Observation and Experiment: An Introduction to Causal Inference}, Cambridge: Harvard University Press. 

\smallskip\setlength{\hangindent}{12pt}
\noindent
Rosenbaum, P. R., and Silber. J. H. (2009), ``Amplification of Sensitivity Analysis in Observational Studies,'' \textit{Journal of the American Statistical Association}, 104, 1398--1405. 

\smallskip\setlength{\hangindent}{12pt}
\noindent
Rubin, D. B. (1974), ``Estimating Causal Effects of Treatments in Randomized and Nonrandomized Studies,'' \textit{Journal of Educational Psychology}, 66, 688--701. 

\smallskip\setlength{\hangindent}{12pt}
\noindent
Samet, J. M., Dominici, F., Curriero, F. C., Coursac, I., and Zeger, S. L. (2000), ``Fine Particulate Air Pollution and Mortality in 20 U.S. Cities, 1987–1994,'' \textit{The New England Journal of Medicine}, 343, 1742--1749.

\smallskip\setlength{\hangindent}{12pt}
\noindent 
Stuart, E. A. (2010), ``Matching Methods for Causal Inference: A Review and a Look Forward,'' \textit{Statistical Science}, 25, 1--21. 

\smallskip\setlength{\hangindent}{12pt}
\noindent
Su, X., Tsai, C. L., Wang, H., Nickerson, D. N., and Li B. (2009), ``Subgroup Analysis via Recursive Partitioning,'' \textit{Journal of Machine Learning Research,} 10, 141--158. 

\smallskip\setlength{\hangindent}{12pt}
\noindent
Wager, S., and Athey, S. (2017), ``Estimation and Inference of Heterogeneous Treatment Effects Using Random Forests,'' \textit{Journal of the American Statistical Association,} (just-accepted)

\smallskip\setlength{\hangindent}{12pt}
\noindent
Zaykin, D. V., Zhivotovsky, L. A., Weestfall, P. H., and Weir, B. S. (2002), ``Truncated Product Method of Combining P-values'' \textit{Genetic Epidemiology}, 22, 170--185.  

\smallskip\setlength{\hangindent}{12pt}
\noindent
Zhang, H. and Singer, B. H. (2010), \textit{Recursive Partitioning and Applications}, \ New York: Springer.

\end{document}